%% file: Nature_MAXIJ0637_Arxix-1.tex
\documentclass[fleqn,10pt]{wlscirep}
\usepackage[utf8]{inputenc}
\usepackage[T1]{fontenc}

\input{z1_Figures.tex}
\input{z2_Tables.tex}

\newcommand{\MAXI}{MAXI~J0637$-$430}
\usepackage{soul}

\include{definitions}

\usepackage[dvipsnames]{xcolor}

\usepackage{xcolor}
\usepackage{comment}
\usepackage{float}
\usepackage[normalem]{ulem}
\usepackage{bm} %properly bolds only variables, while keeping them italicized as required in mathematical notation.

\usepackage{lineno} %enumerate lines
\newcommand{\ion}[2] {#1\,{\sc #2}}
\usepackage{threeparttable}
\title{A self-shadowing black hole accretion disk at sub-Eddington luminosities}

\author[1,2,*]{M.~Armas~Padilla}
\author[1,2]{T.~Muñoz-Darias}
\author[3]{J.~Sánchez-Sierras}
\author[4]{M.~Shidatsu}
\author[1,2]{D.~Mata~Sánchez}
\author[5]{V.~A.~Cúneo}
\author[1,2]{E.~A.~Saavedra}
\author[1,2]{A.~Ambrifi}
\author[6]{M.~C.~Baglio}
\author[1,2]{J.~Casares}
\author[7]{J.~M.~Corral-Santana}
\author[8,9]{R.~P.~Fender}
\author[10]{J.~A.~Fernández-Ontiveros}
\author[6,11,12]{G.~Ponti}
\author[13]{D.~M.~Russell}
\author[1,2]{M.~A.~P.~Torres}
\author[14]{Y.~Ueda}
\author[15,16,17]{F.~M.~Vincentelli}
\author[1,2]{I.~Yanes-Rizo}

\affil[1]{Instituto de Astrofísica de Canarias, E-38205 La Laguna, Tenerife, Spain}
\affil[2]{Departamento de Astrofísica, Universidad de La Laguna, E-38206 La Laguna, Tenerife, Spain}
\affil[3]{Department of Astrophysics/IMAPP, Radboud University, P.O. Box 9010, 6500 GL Nijmegen, The Netherlands}
\affil[4]{Department of Physics, Ehime University, 2-5, Bunkyocho, Matsuyama, Ehime, 790-8577, Japan}
\affil[5]{Leibniz-Institut für Astrophysik Potsdam, An der Sternwarte 16, 14482, Potsdam, Germany}
\affil[6]{INAF–Osservatorio Astronomico di Brera, Via Bianchi 46, I-23807, Merate (LC), Italy}
\affil[7]{European Southern Observatory, Alonso de Córdova 3107, Vitacura, Casilla, 19001, Santiago de Chile, Chile}
\affil[8]{Astrophysics, Department of Physics, University of Oxford, Keble Road, Oxford, OX1 3RH, UK}
\affil[9]{Department of Astronomy, University of Cape Town, Private Bag X3, Rondebosch, 7701, South Africa}
\affil[10]{Centro de Estudios de Física del Cosmos de Aragón, Plaza San Juan 1, E--44001 Teruel, Spain}
\affil[11]{Max-Planck-Institut für extraterrestrische Physik, Giessenbachstraße 1, 85748, Garching bei München, Germany}
\affil[12]{Como Lake Center for Astrophysics, Università degli Studi dell’Insubria, via Valleggio 11, I-22100 Como, Italy}
\affil[13]{Center for Astrophysics and Space Science, New York University Abu Dhabi, PO Box 129188, Abu Dhabi, UAE}
\affil[14]{Department of Astronomy, Kyoto University, Kitashirakawa-Oiwake-cho, Sakyo-ku, Kyoto, Kyoto, 606-8502, Japan}
\affil[15]{Fluid and Complex Systems Centre, Coventry University, CV1 5FB, UK.}
\affil[16]{INAF—Istituto di Astrofisica e Planetologia Spaziali, Via del Fosso del Cavaliere 100, I-00133 Roma, Italy}
\affil[17]{School of Physics \& Astronomy, University of Southampton, Southampton SO17 1BJ, UK}
\affil[*]{m.armaspadilla@iac.es}

\begin{abstract}
\textbf{Accretion onto black holes powers some of the most energetic sources in the Universe, yet its behaviour across the luminosity range remains poorly understood. Specifically, as the luminosity approaches the Eddington limit (\ledd), the inner accretion disk is expected to depart from the standard thin-disk solution, becoming vertically extended. The luminosity threshold and observational signatures of this transition remain, however, uncertain. We present multiwavelength spectroscopy obtained during the decay of a black hole X-ray binary outburst. At the peak ($\sim$0.2~\ledd), a cool optical and near-infrared spectrum dominated by strong recombination lines of hydrogen and \hei\ is observed. However, as the luminosity decreases, it switches to a hot, highly irradiated, nearly featureless spectrum. 
This puzzling behavior reveals the emergence of a vertically extended inner disk at the outburst peak, shielding the outer regions from high-energy radiation produced near the black hole; a scenario supported by contemporaneous X-ray spectra. Observations of similar systems suggest that this phenomenon may be common among black hole accretion disks. Self-shadowing structures at luminosities well below \ledd\ have major implications for black hole accretion flows and their observational properties, affecting inferred black hole spins, X-ray polarization measurements, and the physical conditions of the outer disk and its associated outflows.
}

\end{abstract}

\DeclareUnicodeCharacter{2212}{\ensuremath{-}}
\begin{document}

\flushbottom
\maketitle
\thispagestyle{empty}

\vspace{-0.5 cm}

On November 2, 2019, the Monitor of All-sky X-ray Image \cite{Matsuoka2009} (\textit{MAXI}) reported the detection of an uncatalogued X-ray transient, \MAXI\ \cite{Negoro2019}. Based on its spectral and temporal properties, it was classified as a black hole accreting from a low-mass companion star in an X-ray binary system \cite{Strader2019,Jana2021,Corral-Santana2016}. The black hole outburst was intensively monitored with the X-ray missions Neil Gehrels \swift\ Observatory \cite{Gehrels2004} and Neutron star Interior Composition Explorer (\nicer)\cite{Gendreau2016}. The system was discovered when it was already close to the peak in luminosity, during the disk-dominated soft accretion state\cite{Belloni2011} and was subsequently followed-up throughout its $\sim$6-month-long decay toward quiescence.

Simultaneous optical and near-infrared (OIR) spectroscopy of \MAXI, obtained with the VLT/X-shooter spectrograph\cite{Vernet2011} at three distinct epochs of the outburst decay, reveals a dramatic evolution. The first spectrum, taken at the outburst peak when the system was accreting at $\sim$0.2~\ledd\cite{Soria2022} (see Fig.~\ref{fig:XrayLC_HID} and \hyperref[subsec:evolution]{Methods}), is characterized by multiple emission lines, including strong double-peaked features from hydrogen (Balmer and Paschen series), as well as neutral (\hei) and singly ionized (\heii) helium, also present in an optical spectrum taken with the SOAR telescope one day earlier (see Fig.~\ref{fig:XrayLC_HID} and Extended Data Fig.~\ref{fig:SOAR}). These features are characteristics of active but weakly irradiated accretion disks, such as those observed in disk-dominated accreting white dwarfs\cite{Williams1980}, where X-ray irradiation is minimal. The second and third spectroscopic epochs were obtained two and four weeks later, respectively. Their associated X-ray hardness is similar to that of the first observation, but the X-ray luminosity is significantly lower ($\lesssim$0.1~\ledd; see Fig.~\ref{fig:XrayLC_HID}, \hyperref[subsec:nustar]{Methods}). These spectra reveal a hotter, highly irradiated accretion disk in which the main hydrogen and helium features are virtually absent, showing only weak high-excitation \heii\ emission lines (see Extended Data Fig.~\ref{fig:EW}). This type of optical spectrum persisted as the source declined throughout the soft state, down to a luminosity of at least $\sim$0.04~\ledd\cite{Tetarenko2021}. The dramatic change observed from the first to the second epoch—in particular, the disappearance of most emission lines—cannot be attributed to an increase in the OIR continuum, which, like the X-rays, was observed to drop\cite{Tetarenko2021} (see Extended Data Fig.~\ref{fig:oir_continuum}). Instead, such an abrupt spectral change could be explained by a sudden increase in the ionization parameter of the accretion-disk atmosphere where the OIR emission lines are formed. An increase in ionization would typically result from a corresponding rise in the irradiation level of the outer disk, a behavior therefore expected as luminosity increases (\hyperref[sec:ionization]{Methods}). However, in the case of \MAXI, the opposite is observed. Interestingly, the evolution of the OIR continuum during this phase of the outburst—although likely less sensitive than the emission lines, which originate in the disk atmosphere—also indicates stronger irradiation at lower luminosity\cite{Tetarenko2021}.

While OIR spectroscopy reflects the physical state of the outer disk, the X-ray behavior of the source primarily traces the evolution of the innermost regions of the accretion flow. In this context, X-ray observations of \MAXI\ taken contemporaneously to OIR epochs reveal that standard black hole spectral models, typically applied in disk-dominated states, fail to fit the observed spectra. Along with thermal emission from the disk and thermal Comptonization in the electron corona, an additional soft component is required to accurately reproduce the X-ray continuum (see \hyperref[subsec:xspectramodel]{Methods}). Several scenarios have been proposed to explain this discrepancy. One possibility is that the excess emission originates in the plunging region\cite{Mummery2024}, that is, within the innermost stable circular orbit. It has also been proposed to arise from thermal plasma emission from an ionized outflow, or from a transition between a hotter inner ring near the innermost stable circular orbit and the cooler outer disk\cite{Ma2022}. Alternatively, photons emitted from the inner disk may be bent back toward the disk surface by the strong gravity of the black hole, producing a reflection component through self-irradiation\cite{Cunningham1976, Lazar2021}. This last scenario has also been invoked to explain similar deviations from standard models in other black hole systems\cite{Connors2021, Adegoke2024}, with some requiring the inner disk to become geometrically thick\cite{Zhao2024, Jin2024}. In the case of \MAXI, this interpretation is supported by broadband X-ray analysis based on three epochs, combining \nustar\ with either \nicer\ or \swift\ data, tracing the evolution of the accretion flow during the outburst decay (see \hyperref[subsec:xspectramodel]{Methods}). At the outburst peak, the reflection component is found to be prominent, with the best-fit model attributing $\sim30 \%$ of the total X-ray flux. This is an unusually high fraction that, for moderate black-hole spin values -- as suggested for this source\cite{Soria2022,Mummery2024} -- would require a funnel-like, geometrically thick inner disk with a wide opening angle relative to the equatorial plane\cite{Zhao2024}. Subsequently, as the luminosity fades, the reflection strength gradually decreases, reaching levels consistent with a razor-thin disk geometry\cite{Zhao2024,Connors2020}. The evolution of the disk's temperature profile is also consistent with this scenario (see \hyperref[subsec:xspectramodel]{Methods} and Extended Data Fig.~\ref{fig:threeNustar}). Altogether, a transition from a thick to a standard thin accretion disk provides a unified explanation for both the puzzling OIR and X-ray behavior of the source. In this context, an inner disk structure, developed above $\sim$0.1~\ledd, provides the required X-ray shielding to maintain a low-ionized outer disk during the most luminous phase of the outburst.

Theoretical work suggests that when luminosity approaches \ledd, the standard Shakura–Sunyaev disk solution becomes unstable, giving rise to alternative scenarios. These include the slim\cite{Abramowicz1988} and puffy\cite{Lancova2019, Wielgus2022} disk models -- vertically extended, optically thick configurations expected to form near or above the Eddington limit. Here, we present observational evidence for such vertical inner disk structures. Crucially, this indicates that the self-shadowing geometry is already present well below the Eddington limit, between approximately 10 and 20 per cent of \ledd.

The emergence of inner disk structures has major physical and observational consequences. It modifies the geometry and X-ray emission from the inner flow\cite{Li2010,Wielgus2022,Huang2023}, whose proper modeling is critical for understanding how accretion proceeds close to the black hole and for measuring black hole spin. Several studies have shown that the latter can be significantly overestimated if disk thickness is not accounted for\cite{Riaz2020,Li2010,Wielgus2022,Lancova2023}. Likewise, in thick disks, X-ray polarization levels can reach values up to five times higher than those predicted by standard thin-disk models\cite{West2023,Podgorny2025}, potentially helping to explain the unexpectedly high values recently observed in some soft states\cite{Ratheesh2024}. Finally, inner disk irradiation also affects the outer regions, where OIR lines are formed. In addition to the disk atmosphere, several studies have shown that these features can also trace the presence of powerful disk winds\cite{Munoz-Darias2026}. These outflows, initially discovered in X-rays\cite{Ponti2012}, carry significant mass and angular momentum\cite{Munoz-Darias2016,Sanchez-Sierras2023b,Tetarenko2018}. However, their detection in the OIR domain remains challenging due to overionization effects. Shielding by high-density clumps and geometrical self-shadowing—such as that inferred in this study—are the favored mechanisms to preserve low-ionization conditions in parts of the wind\cite{Motta2017b,Munoz-Darias2022}.

Possible observational signatures of evolving inner accretion flow geometry during disk-dominated states have been explored primarily through X-ray spectral modeling\cite{Kubota2004,Zhao2024}. Our observations of \MAXI\ reveal that changes in the inner flow geometry have a direct, measurable impact on the irradiated outer disk. This provides multiwavelength evidence connecting the evolution of both the inner and outer regions of a black hole accretion disk. One key property might have been fundamental in making \MAXI\ such a revealing system. Several observational features point to a short orbital period of only a few hours ($\lesssim$5~hours)\cite{Yanes-Rizo2026,Tetarenko2021,Ma2022,Soria2022}, implying that its accretion disk is smaller than those of most black hole transients\cite{Corral-Santana2016}. While the size of the X-ray–emitting region does not depend on the full disk extent, the OIR–emitting region is significantly more compact in short-period systems. As a result, the physical state of their outer disks is expected to be relatively homogeneous and, consequently, more prone to abrupt variations as soon as the irradiation conditions change. 

Direct comparison with the shortest orbital period black holes ($\sim$ 2 to 6 hours) is challenging, primarily due to the lack of OIR spectroscopy during disk-dominated states. As such, \MAXI\ represents a unique case. Nevertheless, featureless spectra, consistent with a highly irradiated disk, seem to be common in these systems (see \hyperref[sec:OIRshortPo]{Methods}). The proposed scenario can be tested by considering the few systems with known orbital period and high-quality spectroscopy, in particular H$\alpha$ observations, obtained at different luminosities throughout the soft state (see Fig.~\ref{fig:HalphaEvol}). The closest case to \MAXI\ is that of MAXI~J1803$-$298 (orbital period $\sim$7 hours). Although its behavior is not as extreme, it exhibited a significant drop in line intensity during the disk-dominated state as the X-ray luminosity decayed \cite{MataSanchez2022}. The H$\alpha$ equivalent width evolved from 2.5\AA\ to $\sim$0.9\AA\ as the luminosity dropped from $\sim$0.3~\ledd\ to $\sim$0.06~\ledd, reemerging again at $<$0.03~$L_{\mathrm{Edd}}$. Gradually moving toward longer orbital periods, Swift~J1727.8$-$1613 (10 hours orbital period) shows a milder version of this effect, with line intensity decreasing as the luminosity dropped from $\sim$0.3~\ledd\ to $\sim$0.1~\ledd \cite{MataSanchez2024} (Fig.~\ref{fig:HalphaEvol}). Conversely, MAXI~J1820$+$070 (a 16-hour orbital period black hole) exhibits strong emission lines characteristic of relatively low-irradiated accretion disks throughout its several-month-long disk-dominated state \cite{Munoz-Darias2019, Sanchez-Sierras2020}(Fig.~\ref{fig:HalphaEvol}). Additionally, GX~339$-$4 (42 hours) showed strong low-ionization lines during the luminous peak of its 2021 soft state\cite{Ambrifi2025}, and similar behavior was observed in the optical and infrared emission lines of 4U~1543$-$47 (26 hours) while accreting in an ultra-luminous state, likely around the Eddington limit\cite{Sanchez-Sierras2023}. Interestingly, during disk-dominated states, the X-ray spectra of most of these sources also deviate from standard spectral models, resembling the behavior seen in \MAXI\cite{Adegoke2024,Zhao2024,Fabian2020}. Our results imply the presence of self-shadowing effects already at moderate luminosities, although their impact on the outer disk only becomes apparent in systems with relatively small disks (orbital period $\lesssim$11--15 hours). These effects may explain why low-ionization H and He emission lines are consistently observed in some luminous systems, even when accreting near the Eddington limit: the inflated inner disk provides shielding at the highest accretion rates. And as the X-ray irradiation decays, the distant, cooler outer regions in long-period binaries continue to emit lines even after the shielding disappears.

Figure~\ref{fig:sketch} (top panel) illustrates how the disk geometry would evolve throughout the outburst decay during the disk-dominated state in a physically small system such as \MAXI. At the peak of the outburst, when the system is accreting above $\sim10$~\%~\ledd, the inner regions of the accretion disk deviate from the thin geometry and become vertically extended, likely due to radiation pressure. This newly emerged structure shields the outer regions of the disk from intense X-ray irradiation, resulting in a lower ionization state. For instance, an illustrative vertical thickness of merely five per cent of the radius (i.e. $H/R \gtrsim 0.05$) would suffice to shield the outer disk in systems with orbital periods of tens of hours (\hyperref[sec:H/R]{Methods}). As a result, strong recombination lines, such as the double-peaked lines of H and \hei, can be emitted, which would otherwise be suppressed. As the X-ray luminosity decays and the radiation pressure decreases, the accretion disk returns to the thin Shakura–Sunyaev solution\cite{Shakura1973}. Consequently, the outer regions are exposed to severe X-ray irradiation, leading to strong ionization, which renders the OIR spectrum featureless, with only a weak presence of high-excitation emission lines (e.g. \heii). As the luminosity decays further and the X-ray irradiation decreases sufficiently, the outer regions cool down, allowing the emission lines to re-emerge, as generally observed in black hole accretion disks at low luminosities (e.g. during quiescence\cite{Casares2015}). The evolution of the inner disk is expected to be similar in long-period systems ($\gtrsim11$–15~hours). However, the state of the outer regions responsible for the OIR emission can differ, owing to their larger disk radii. As illustrated in Fig.~\ref{fig:sketch} (bottom panel), the previous scenario remains applicable, but with a key difference: despite the high X-ray irradiation, the outer regions are located much farther from the black hole than in short-period systems. As a result, in the absence of self-obscuration (i.e. below 0.1~\ledd), even though the inner regions are fully ionized and do not emit lines, the ionization remains low enough at large radii to allow the formation of low-ionization emission lines (\hyperref[sec:ionization]{Methods}). 

The present study of \MAXI\ provides strong evidence for a self-shadowing inner disk structure in an accreting black hole. Theoretical work has explored how the geometry of the inner accretion flow deviates from the standard thin-disk configuration when the accretion rate approaches or exceeds the Eddington limit\cite{Abramowicz1988}. Our results show that the disk departs from the classic thin solution already at $\sim0.1$--$0.2$~\ledd, and that this has a strong impact on the physical conditions of the outer disk. Understanding inner disk geometry is essential for spectral modeling in the strong-gravity regime and, ultimately, for measuring black hole spin\cite{Li2010, Wielgus2022, Zdziarski2025}. In addition, it is crucial for interpreting spectral variability, polarization measurements, and the formation of disk winds across a wide range of accreting black holes---from Galactic transients to active galactic nuclei. MAXI~J0637$-$430 thus offers a rare and revealing view into how structural changes in the accretion flow shape the observable signatures of black hole accretion.
\newpage

\FigureXrayLCHID

\FigureHalphaPorb

\clearpage
\FigureSketchver

\newpage
%\bibliography{references}

\section*{Methods}

\section*{Observations and data reduction} \label{sec:obs}
\subsection*{Outburst monitoring: \swift\ and \nicer}
The $\sim$6~month 2019--2020 outburst of \MAXI\ was monitored extensively with the \textit{Neil Gehrels} \swift\ Observatory \cite{Gehrels2004} and the Neutron star Interior Composition Explorer (\nicer)\cite{Gendreau2016}. Observations with the X-ray Telescope \cite{Burrows2005} on board \swift\ were carried out between 2019 November 7 and 2020 March 30 (ObsIDs 00012172001 – 00012172096). Most of the observations were performed in Windowed Timing (WT) mode, except the last 16 pointings that were taken in Photon Counting (PC) mode. We excluded from our analysis observations shorter than 200~sec and those WT observations in which part of the source region was located outside the edges of the WT window. We processed the data using \textsc{HEASOFT} v.6.35 software. Data reduction was carried out by running the {\tt xrtpipeline} task, in which standard event grades of 0–12 and 0–2 were selected for the PC and WT mode observations, respectively. For every observation, spectra, light curves and images were obtained using \textsc{xselect}. To extract the source events, we used a circular region of 30 pixels
($\sim$ 70 arcsec) radius centered at the source position. The first half of the observations were affected by pile-up, which we mitigated by excising the innermost 4–8 pixels from the extraction region. For the WT-mode observations, background events were extracted from an annular region centered on the source, with inner and outer radii of 70 and 130 pixels, respectively. For the PC-mode data, background spectra were extracted from three circular regions of similar size located in source-free areas of the sky. We created exposure maps and ancillary and response files following the standard \swift\ analysis threads. The latest versions of the response matrix files were taken from the High Energy Astrophysics Science Archive Research Center (HEASARC) calibration database (CALDB). Finally, we grouped the spectra to have a minimum of 20 photons per bin.

\nicer, onboard the International Space Station, monitored the source between 2019 November 3 and 2020 April 9 (ObsIDs 2200950101–89 and 3200950101–12). Data were processed using the \textsc{nicerdas} v.12 using the latest calibration files from CALDB applying standard screening criteria to produce cleaned event files (we excluded data with a pointing offset of $>$54 arcsec, bright Earth limb angle $<$30$^{\circ}$, dark Earth limb angle $<$15$^{\circ}$, and in the South Atlantic Anomaly). Background spectra were generated using the \texttt{nibackgen3C50} tool \cite{Remillard2022}, which models the instrumental background based on space weather proxies. Source spectra were extracted in the 0.5–10~keV energy range and corrected for energy oversampling. Finally, all spectra were grouped to ensure a minimum of 30 counts per bin.

\subsection*{\nustar\ data }
\phantomsection
\label{subsec:nustar}

\MAXI\ was observed with the Nuclear Spectroscopic Telescope Array\cite{Harrison2013} (\nustar) in 8 epochs. The first three \nustar\ observations were taken on November 5, 13, and 25, 2019, covering the peak of the outburst (OBsIDs 80502324002, 80502324004 and 80502324006, respectively). The first observation is simultaneous with the first VLT epoch (OIR-1), while the second and third are contemporaneous with the period spanned by the remaining VLT observations (i.e. immediately before and after epoch OIR-2, and prior to OIR-3; see top panel of Extended Data Fig.\ref{fig:threeNustar}). We used \nustar\ Data Analysis Software (NUSTARDAS) contained in the HEASoft v.6.35 package to reduce these data. We processed the data using the {\tt\string nupipeline} task, and extracted the spectra, light curves, and response files from both focal plane modules A (FPMA) and B (FPMB) with the {\tt\string nuproducts} tool. For the source, we used a circular region of 80~arcsec radius, while for the background, we used a circular region of 120~arcsec radius located in a source-free region of the same detector quadrant. We grouped the spectral data into bins containing a minimum of 20 photons.

\subsection*{Optical and near-IR spectroscopy}\label{subsec:OIR}
We observed 3 epochs of optical and near-infrared spectroscopy that were carried out with the X-shooter spectrograph \cite{Vernet2011}, located in the Very Large Telescope (Paranal, Chile). X-shooter is a wide-band spectrograph covering from 3000 to 24800~\AA~in 3 arms, which work simultaneously: UVB arm (3000--5560\AA), VIS arm (5500--10200\AA) and NIR arm (10200--24800\AA). Two epochs were observed on the nights of 5 and 17–18 November, while the third was taken on the night of 2 December. For each epoch, we took 4-8 exposures  with total exposure times ranging from 2040 to 2400 sec. The instrument was set in the long slit configuration with a slit-width of 1.0 arcsec (UVB) and 0.9 arcsec (VIS/NIR) with a 2x2 binning, yielding resolutions of 55~km~s$^{-1}$ (UVB), 34~km~s$^{-1}$ (VIS) and 54~km~s$^{-1}$ (NIR). The data were reduced with ESO X-shooter pipeline v.3.5.0. Biases were subtracted via overscan and flatfield correction was applied. The pipeline computed a preliminary physical model before a precise spatial and wavelength calibration with the geometry solution in order to obtain the 2D calibrated and sky-subtracted spectra. The images were corrected from instrument flexures with attached flexure compensation frames taken just before each observation. Cosmic rays were removed with the Laplacian edge detection algorithm\cite{VanDokkum2001}. Finally, the spectra were extracted and flux calibrated with a standard flux star observed the same night of each epoch.
All spectra were carefully corrected from telluric absorptions with \textsc{Molecfit} v1.5.9 fitting the telluric features to a computed atmosphere model and doing fit loops in order to tune molecular density values. The data analysis was carried out with \textsc{Molly} tool and with custom \textsc{Python} 3.7 routines. The full, normalized optical and near-infrared spectra are presented in Extended Data Figs.~\ref{fig:VLT_1_2} and \ref{fig:VLT_1_3}.

An additional epoch of optical spectroscopy was obtained on 4 November 2019, one day before our first VLT epoch (OIR-1). The data were retrieved from the public archive\cite{Strader2019} (NOIRLab Astro Data Archive; https://astroarchive.noirlab.edu). The observations were carried out with the Goodman Blue and Red Camera instruments mounted on the 4.1-m SOAR Telescope at Cerro Pachón, Chile\cite{Clemens2004}. The total exposure time was 1200~sec, covering a combined wavelength range of 3900–7800\AA\ at a spectral resolution of $\sim$300~km~s$^{-1}$. The spectrum was reduced using standard IRAF and \textsc{Molly} routines, including bias subtraction, flat-field correction, and wavelength calibration. The final reduced spectrum is shown in Extended Data Fig.~\ref{fig:SOAR}.

\section*{Evolution of the optical emission lines}

To quantify the spectral evolution across the OIR-1 to OIR-3 phases, we measured the equivalent width (EW) of the main optical emission lines (\ha, \hei\ $\lambda$5876, and \heii\ $\lambda$4686). We exclude H$\beta$ from this analysis because weak emission is difficult to disentangle from broad underlying Balmer absorption \cite{MataSanchez2026}; the H$\beta$ region is also commonly affected by diffuse interstellar bands (DIBs). As an additional consistency check at the earliest epoch, we used the SOAR spectrum obtained one day prior to OIR-1. Although the spectral resolution of this spectrum is lower than that of the VLT/X-shooter data, the emission lines are clearly detected and their EWs are consistent with those measured during OIR-1 (Extended Data Fig.~\ref{fig:SOAR}).

The temporal evolution of the equivalent widths is shown in Extended Data Fig.~\ref{fig:EW}. These measurements reveal a clear progressive decrease in line strength after OIR-1, with the Balmer and \hei\ lines fading to nearly undetectable levels by OIR-3. \heii\ $\lambda$4686 also declines, but remains visible throughout, indicating a much hotter, ionised, irradiated accretion disk.

\section*{X-ray spectral analysis}
\phantomsection
\label{sec:xanalysis}
We used \textsc{xspec} (v.12.15)\cite{Arnaud1996} to analyse the X-ray spectra. In order to account for interstellar absorption, we used the Tuebingen-Boulder Interstellar Medium absorption model ({\tt TBabs} in \textsc{xspec}) using standard cross-sections\cite{Verner1996} and abundances\cite{Wilms2000}. We fixed the hydrogen column density (\Nh) to \Nh = 4$\times$10$^{20}$\nh\ in all our model fits, based on the interstellar absorption bands identified in our optical spectra. Specifically, we used the correlation between $E(B-V)$ and the strongest component of the \ion{Na}{i} doublet (NaD1, 5890\AA)\cite{Munari1997}. From epoch OIR-1, which has the highest signal-to-noise ratio among our observations, we measured an equivalent width of $EW = 0.176 \pm 0.010$~\AA. This corresponds to a color excess of $E(B-V) = 0.05$. Using the standard relation $A_{\rm V}$ = 3.1$E(B-V)$ and applying the canonical \Nh–$A_{\rm V}$ conversion\cite{Predehl1995}, we obtained \Nh~$\sim$4$\times$10$^{20}$\nh, consistent with previously reported values\cite{Tetarenko2021}.

\subsubsection*{Outburst evolution, peak luminosity and distance estimates}
\phantomsection
\label{subsec:evolution}
We used a combination of multicolour disk and Comptonization components ({\tt diskbb+Nthcomp} in \textsc{xspec}) to fit all \swift\ and \nicer\ spectra. We left free all the parameters. We note that for several observations the model yields results that are not physically acceptable, especially for the soft state observations. We are interested in obtaining statistically good fits (\chired $\lesssim$1.2) in order to calculate unabsorbed fluxes to build the Hardness-Intensity Diagram (HID; see Fig. \ref{fig:XrayLC_HID}). We defined the hardness as the ratio of unabsorbed flux on the 4–10 keV and 1–4 keV bands, respectively.

The spectral and timing properties indicate that the source was already in the disk-dominated (i.e. soft) phases during the earliest recorded epochs\cite{Lazar2021, Jana2021}. It remained in this state, with luminosity gradually decreasing, for the first $\sim$70 days of the outburst, after which it transitioned to the hard state (see Fig.~\ref{fig:XrayLC_HID}). In black hole transients, this soft-to-hard transition is typically observed at 0.01--0.04~\ledd, with a mean value of $\sim$0.02~\ledd\cite{Maccarone2003}. For \MAXI, this implies a peak luminosity in the range of $\sim$10--30\%~\ledd. In agreement with previous work\cite{Soria2022}, we adopt a peak luminosity of 0.2~\ledd, inferred from a soft-to-hard transition occurring at $\sim$0.02~\ledd.

The above results also allow us to estimate the distance to the source. The transition occurred at an unabsorbed 1--10~keV flux of $F_{\mathrm{X}} = 4.24 \times 10^{-10}~\flux$, corresponding to a bolometric flux of $F_{\mathrm{bol}} = 1.2 \times 10^{-9}~\flux$. Assuming a soft-to-hard transition at 0.02~\ledd, we obtain a distance of 12.6~kpc (8.8~kpc and 17.8~kpc for transition luminosities of 0.01 and 0.04~\ledd, respectively), consistent with previous estimates\cite{Jana2021,Soria2022,Abdulghani2024}. At 12.6~kpc, \MAXI\ would lie at a Galactic height of $|z| \sim 4.5$~kpc and a Galactocentric distance of $\sim$17~kpc, placing the system in the outskirts of the Galaxy. All known black hole transients are located at $|z| < 4.5$~kpc\cite{Corral-Santana2016}, and thus distances beyond $\sim$13~kpc would be difficult to reconcile. This further supports the inferred 0.2~\ledd\ peak luminosity, which would in any case not exceed $\sim$0.3~\ledd, that is, well below the near-Eddington regime. For comparison, a peak luminosity above $\gtrsim$0.5~\ledd\ would imply (in addition to an exceptionally high soft-to-hard transition luminosity) a distance greater than 23~kpc, a Galactocentric distance $>$26~kpc, and a height $|z| > 8$~kpc---far beyond the observed distribution of Galactic black hole transients. Distances below $\sim$8~kpc \cite{Ma2022}, on the other hand, would imply a soft-to-hard transition below $\sim$0.01~\ledd, significantly lower than typically observed in black hole transients, and a peak outburst luminosity below $\sim$0.1~\ledd. In this case, the system would enter a regime where disk self-obscuration occurs at even lower luminosities, making the proposed scenario more extreme. 

\subsubsection*{Broad band X-ray spectroscopy}
\phantomsection
\label{subsec:xspectramodel}

Three broad-band X-ray epochs trace the decay of the outburst and span the same period as our OIR VLT observations. The first epoch, combining \nicer\ and \nustar\ data, was simultaneous with epoch OIR-1 on 5 November 2019. The second and third X-ray epochs, based on \swift+\nustar\ observations taken on 13 and 25 November 2019, occurred immediately before and after epoch OIR-2, and prior to OIR-3, respectively (see the top panel of Extended Data Fig.\ref{fig:threeNustar}). We performed joint fits of the 1–10~keV \nicer\ (or \swift) and 5–70~keV \nustar\ spectra, tying the spectral parameters across the three detectors. To account for cross-calibration uncertainties, we included a multiplicative constant component (\texttt{constant}), fixed to 1 for the \nicer\ (or \swift) spectrum and left free to vary for the \nustar\ cameras (FPMA and FPMB). 

As a first step, we attempted to fit the spectra using the standard two-component model typically applied to black hole soft states. This includes a soft thermal component that accounts for  emission from the accretion disk combined with a hard component modelling the inverse-Comptonized emission from the corona. Thus, we used a multicolor disk plus a thermally Comptonized continuum (\texttt{diskbb+Nthcomp} in \textsc{xspec}; \cite{Mitsuda1984,Zdziarski1996,Zycki1999}). A disk geometry was assumed for the seed photons, with their temperature ($kT_{\rm seed}$) tied to the inner disk temperature ($kT_{\rm diskbb}$). The model failed to provide an acceptable fit for any of the three datasets, with all residuals showing clear structures (Extended Data Fig.\ref{fig:nustarspectra}). The corresponding fit statistics were $\chi^2$/d.o.f. = 3668.5/1266, 2684.5/1842, and 1545.8/1137 for the first, second, and third epoch, respectively. Replacing \texttt{diskbb} with either \texttt{kerrbb} \cite{Li2005} or the \textit{p}-free radial temperature-profile disk model \texttt{diskpbb} \cite{Mineshige1994,Kubota2005} did not improve the fit. Substituting the Comptonization component with the \texttt{relxillCp} model \cite{Garcia2014}, which simultaneously accounts for Comptonization and reflection, provided an improvement, but the fit remained statistically unacceptable (p-value $< 10^{-33}$), with significant residuals still visible \cite{Lazar2021} (see, e.g., Extended Data Fig.\ref{fig:nustarspectra} for the first broad-band X-ray epoch). 

The best fits were obtained by adopting a reflection model in which the incident spectrum is a blackbody, as previously proposed for this source \cite{Lazar2021}, specifically \texttt{relxillNS} \cite{Garcia2022}, which, despite assuming a simplified single-temperature blackbody for the illuminating spectrum, is widely used as a proxy for the reflection emission expected from returning disk radiation in black hole soft states\cite{Connors2020,Connors2021,Lazar2021,Wang2021a,Zhao2024}. To obtain a more physically consistent evolution of the disk parameters, we adopt the \texttt{diskpbb} model \cite{Mineshige1994,Kubota2004}, in which the radial temperature profile follows $T(r)\propto r^{-p}$, with the index $p$ treated as a free parameter. The standard thin-disk solution is recovered for $p=0.75$, while lower values allow for departures from the classical thin-disk configuration. Thus, our final model was \texttt{diskpbb+Nthcomp+relxillNS}. The emissivity indices ($q_{\rm in}$ and $q_{\rm out}$)  of \texttt{relxillNS} were fixed to 3, consistent with the shallower emissivity profiles expected from disk self-irradiation \cite{Wilkins2020}. The inner radius ($R_{\rm in}$) was set at the ISCO, as commonly assumed for accretion disks in the soft state, and the redshift to zero \cite{Connors2021,Lazar2021}. The reflection fraction was fixed at $R_{\mathrm{f}}=-1$, ensuring that only the blackbody-reflected component is included \cite{Zhao2024}. The iron abundance ($A_{\rm Fe}$), inclination angle ($i$), and spin ($a_{*}$) were tied across the three epochs. The values derived for these parameters are $a_{*} \sim 0.58$, $i \sim 69^\circ$, and $A_{\rm Fe} \sim 1.4$, broadly consistent with those reported in previous studies\cite{Lazar2021,Soria2022,Mummery2024}. The remaining parameters were left free. The best-fitting results and corresponding fluxes from this analysis are presented in Extended Data Table~\ref{tab:fit_three_epochs_Nustar} and Extended Data Fig.~\ref{fig:threeNustar}.

The evolution of the reflection contribution to the total X-ray flux reveals a clear and significant trend (green dashed line in the fitted spectra of Extended Data Fig.~\ref{fig:threeNustar}). Over time, the relative strength of the reflection component steadily decreases, indicating a diminishing contribution of reprocessed emission as the outburst decays. This trend is consistent with the idea that the geometry of the inner accretion flow evolves significantly throughout the event. The reflection component accounts for approximately 30\% of the total flux during epoch OIR-1. Such high values cannot be produced by a standard, razor-thin disk, which yields at most $\sim$10-15\% of returning photons even for moderately high black-hole spin values ($a_{*} \lesssim 0.8$)\cite{Wilkins2020,Connors2020,Dauser2022}, as is likely the case for \MAXI\cite{Soria2022, Mummery2024}. General relativistic ray-tracing simulations show that such high fractions require a funnel-like thick disk geometry, where the inner disk becomes geometrically thick with a wide opening angle relative to the equatorial plane \cite{Zhao2024}. The contribution of the reflection component decreases to $\sim$18\% in the second broad-band X-ray spectrum (immediately before epoch OIR-2), and drops to a very low level ($\sim$4\%) by the third epoch (after epoch OIR-2 and prior to epoch OIR-3), consistent with the fraction of photons expected to return to the disk under the assumption of a standard thin geometry\cite{Connors2021,Zhao2024}.
The radial temperature index of \texttt{diskpbb} also suggests a departure from the standard thin-disk regime during the first observation. While X-2 and X-3 yield $p \sim 0.75$, consistent with a standard disk configuration, X-1 instead shows a lower value, $p \sim 0.58$, often associated with a geometrically thickened inner disk \cite{Isobe2012,Soria2015}. This evolution is consistent with the scenario proposed to account for the optical and near-infrared behavior, in which the inner disk was initially inflated during the luminous phase of the outburst, producing self-shielding of the outer disk, before collapsing back into a geometrically thin configuration as the source luminosity decayed.

Given the limitations of the adopted spectral models and the uncertainties in key system parameters (e.g. inclination and spin), these results should be interpreted qualitatively rather than quantitatively, with the interpretation driven by the overall behavior and its evolution across epochs, rather than by the exact values of the fitted parameters in any individual spectrum. A quantitative treatment would require more sophisticated physical reflection models, including self-consistent treatment of returning radiation, light bending, higher-order reflection, and the detailed angular and thermal structure of the illuminating spectrum \cite{Dauser2022,Mirzaev2024}.

Alternative explanations for the observed excess soft X-ray emission include high-density reflection\cite{Tomsick2018,Jiang2019,Connors2021}, warm Comptonization\cite{Petrucci2018,Papavasileiou2026}, or different coronal geometries, as well as those previously proposed for \MAXI\ (e.g., plunging region\cite{Mummery2024} or ionized outflow\cite{Ma2022}). However, none of these scenarios readily explains the optical and infrared behavior, in particular the disappearance of recombination lines as the luminosity drops.

\section*{Self-shadowing of the outer accretion disk}
\phantomsection
\label{sec:H/R}
To estimate the radial extent of the shadowed region produced by a vertically extended inner flow, we adopt the standard prescription for irradiated accretion disks\cite{Cunningham1976,Frank2002,Tomaru2019}, in which the disk thickness increases with radius as $H/R \propto R^{2/7}$.
This scaling can be expressed as
\[
H/R \approx 1.5\times10^{-3} \left(\frac{L}{L_{\rm Edd}}\right)^{1/7}
\left(\frac{M}{M_\odot}\right)^{-1/7}
\left(\frac{R}{R_g}\right)^{2/7}
\]

where $L/L_{\rm Edd}$ is the Eddington-scaled luminosity, $M$ the black hole mass in units of $M_\odot$, and $R$ the radius in units of the gravitational radius $R_g$.

In our case ($M \approx 8 M_\odot$ and $L \approx 0.2~L_{\rm Edd}$) this becomes
\[
H/R \approx 8.9\times10^{-4}\left(\frac{R}{R_g}\right)^{2/7}
\]
For the outer disk to remain within the geometrical shadow of the inner structure, i.e. $(H/R){\rm inner} > (H/R){\rm outer}$, the maximum outer radius is
\[
R_{\rm out} \approx \left[\frac{(H/R)_{\rm inner}}{8.9\times10^{-4}}\right]^{7/2} R_g
\]
For an inner disk thickness of $H/R \approx 0.05$, this yields $R_{\rm out} \approx 1.3\times10^{6} R_g$, which, using the standard scaling $R_{\rm out} \propto P_{\rm orb}^{2/3}$, corresponds to orbital periods of the order of several hundred hours. This is larger than the value implied by the standard thin-disk solution ($H/R \approx 0.001 - 0.01$), but significantly smaller than that expected near \ledd\ ($H/R \gtrsim 0.2$)\cite{Shakura1973, Chen2004,Sadowski2011}.

\section*{Ionization parameter}
\phantomsection
\label{sec:ionization}
Our OIR spectra reveal significant changes between epochs OIR-1 and OIR-2 (Fig. \ref{fig:XrayLC_HID}). For the hydrogen recombination lines to disappear, the ionization parameter, defined as

\[
\xi = \frac{L}{n\,r^2},
\]

must increase by at least one order of magnitude\cite{Kallman1982}. Assuming, for simplicity, that the lines originate from the same region of the disk and that the density remains constant (i.e. $r$ and $n = \text{const.}$), this rise in $\xi$ would require a corresponding increase in luminosity ($L$). However, the opposite behavior is observed. If the luminosity and density instead remain constant, a decrease in $\xi$ by one order of magnitude would require the emission region to move three to four times farther from the X-ray source. In this context, the size of the accretion disk plays a key role. For a black hole with $M = 8~M_{\odot}$, the outer disk radius\cite{Frank1992} is estimated to be $R_{\mathrm{out}} \approx 9.4 \times 10^5$~km (or $\sim 7.9 \times 10^4\,R_{\rm g}$) for a system with an orbital period of five hours. In contrast, for a 17-hour orbital period, the disk can extend to $R_{\mathrm{out}} \approx 2.1 \times 10^6$~km (or $\sim 1.8 \times 10^5\,R_{\rm g}$).

These considerations are consistent with a scenario in which long-period systems provide the spatial scales necessary to accommodate lower-ionized material without requiring a significant drop in luminosity or density (see Fig.~\ref{fig:HalphaEvol}). In short-period systems, the smaller disk size limits the range of radii over which such a drop in $\xi$ can occur, favoring the observation of a hot, highly ionized, featureless optical spectrum under strong irradiation.

\smallskip 
\section*{Optical and infrared spectra of short-period X-ray binaries}
\phantomsection
\label{sec:OIRshortPo}

Direct comparison with short-period black holes is challenging, primarily because they have only been observed briefly in the OIR during their outbursts, with limited follow-up to track their spectral evolution and no observations available during the soft state. This is the case for the black hole X-ray binary Swift~J1753.5$-$0127, which, with a 3.4-hour orbital period, has shown only hard-state outbursts. The few optical spectra obtained during these events reveal a featureless continuum at the most luminous levels\cite{CadolleBel2007, Durant2009, Neustroev2024}, with disk emission lines detected only at low luminosities during the rise toward outburst\cite{Torres2005}, in the low-luminosity hard state ($<$0.001~\ledd)\cite{Rahoui2015, Neustroev2014}. Similarly, the 2.8-hour system Swift~J1357.2$-$0933 shows a highly irradiated, hot accretion disk, where only very weak emission lines are present at the outburst peak\cite{Corral-Santana2013, Jimenez-Ibarra2019b, Charles2019}, emerging more strongly as the system transitions to quiescence\cite{MataSanchez2015a}. The shortest-period black hole known so far, the 2.4-hour MAXI~J1659$-$152, also shows emission lines in its OIR spectra, but only during the very early stages of the outburst, when it was still in the hard state at $<$0.01~\ledd\cite{Kaur2012a}. Finally, Swift~J1910.2$-$0546, with an orbital period between 2.3 and 4.5 hours, displayed an OIR spectrum dominated by continuum emission, lacking Balmer and \hei\ lines and showing only weak \heii\ emission at 0.05–0.1~\ledd, with H$\alpha$ detected only at $<$0.001~\ledd\cite{Charles2012, Corral-Santana2025}.

\subsection*{Correspondence and requests for materials} should be addressed to Montserrat Armas Padilla

\section*{Acknowledgements}

We acknowledge support by the Agencia Estatal de Investigación via the Plan de Generación de Conocimiento through grants PID2021-124879NB-I00, PID2022-143331NB-100 and PID2024-161863NB-I00. MAP and DMS also acknowledge support through the Ramón y Cajal grants RYC2022-035388-I and RYC2023-044941, funded by MCIU/AEI/10.13039/501100011033 and FSE+. VAC acknowledges support from the Deutsches Zentrum für Luft- und Raumfahrt (DLR) under contract 50 OR 2405. JAFO acknowledges financial support from MICIU/AEI/10.13039/501100011033575, the European Union NextGenerationEU/PRTR, and ERDF/EU through grants CNS2023-145339 and PID2023-147386NB-I00. Y.U. acknowledges the support from the Kyoto University Foundation. JSS is supported by the European Union (ERC, StarStruck, 101095973; PI Jonker). R.F. acknowledges support from the UKRI, The Hintze Family Charitable Foundation and European Research Council Synergy Grant ’Blackholistic’ (grant agreement number 101071643). M.C.B. acknowledges support from the INAF Fundamental Research Grant XBOOM. M.S. acknowledges support by Grants-in-Aid for Scientific Research 23K03459 and 24H01812 from the Ministry of Education, Culture, Sports, Science and Technology (MEXT) of Japan. D.M.R. is supported by Tamkeen under the NYU Abu Dhabi Research Institute grant CASS. G.P. acknowledges financial support from the European Research Council (ERC) under the European Union’s Horizon 2020 research and innovation program HotMilk (grant agreement No. 865637) and from the Framework per l’Attrazione e il Rafforzamento delle Eccellenze (FARE) per la ricerca in Italia (R20L5S39T9). Based on observations collected at the European Southern Observatory under ESO programme 0104.D-0251(A). This research has made use of data from the Neil Gehrels Swift Observatory, the Neutron star Interior Composition Explorer (NICER), and the Nuclear Spectroscopic Telescope Array (NuSTAR), obtained from the High Energy Astrophysics Science Archive Research Center (HEASARC), a service of the Astrophysics Science Division at NASA/GSFC.

\section*{Author contributions statement}
M.A.P. led the article, developed the main physical interpretation, performed the X-ray calibration and analysis and led the writing. T.M.D. contributed to developing the main physical interpretation, obtained and coordinated the VLT/X-shooter observations, and contributed to the writing. J.S.S. performed the VLT/X-shooter calibration and reduction. D.M.S. performed the SOAR calibration and reduction, and contributed to the OIR analysis. M.S. contributed to the broad-band analysis and irradiation interpretation. V.A.C. contributed to the NICER calibration. M.C.B. provided photometric data of the source. I.Y.R. contributed to the photometric analysis. J.M.C.S. contributed to coordinate the VLT/X-shooter observations. J.C., R.P.F., J.A.F.O., G.P., D.M.R., M.A.P.T. and Y.U. contributed to the interpretation of the results and reviewed the manuscript. E.A.S., A.A. and F.M.V. contributed to discussions of the results. All authors discussed the results and contributed to the article.

\subsection*{Competing interests}
The authors declare no competing interests.

\renewcommand{\tablename}{Extended Data Table}
\setcounter{table}{0}
\newpage
\TableFitNuSTARNEW
\newpage

\renewcommand{\figurename}{Extended Data Figure}
\setcounter{figure}{0}

\FigureOIRCont
\ThreeNustar
\FigureNUSTAR
\FigureEW
\FigureVLTUnoDos
\FigureVLTUnoTres
\FigureSOAR

\clearpage

\input{Nature_MAXIJ0637_Arxix-1.bbl}
\end{document}

%% file: z1_Figures.tex
\newcommand{\FigureXrayLCHID}{
\begin{figure}[H]
\centering

\includegraphics[keepaspectratio,width=1.0\columnwidth, trim=0.6cm 0.0cm 0.0cm 0.0cm]{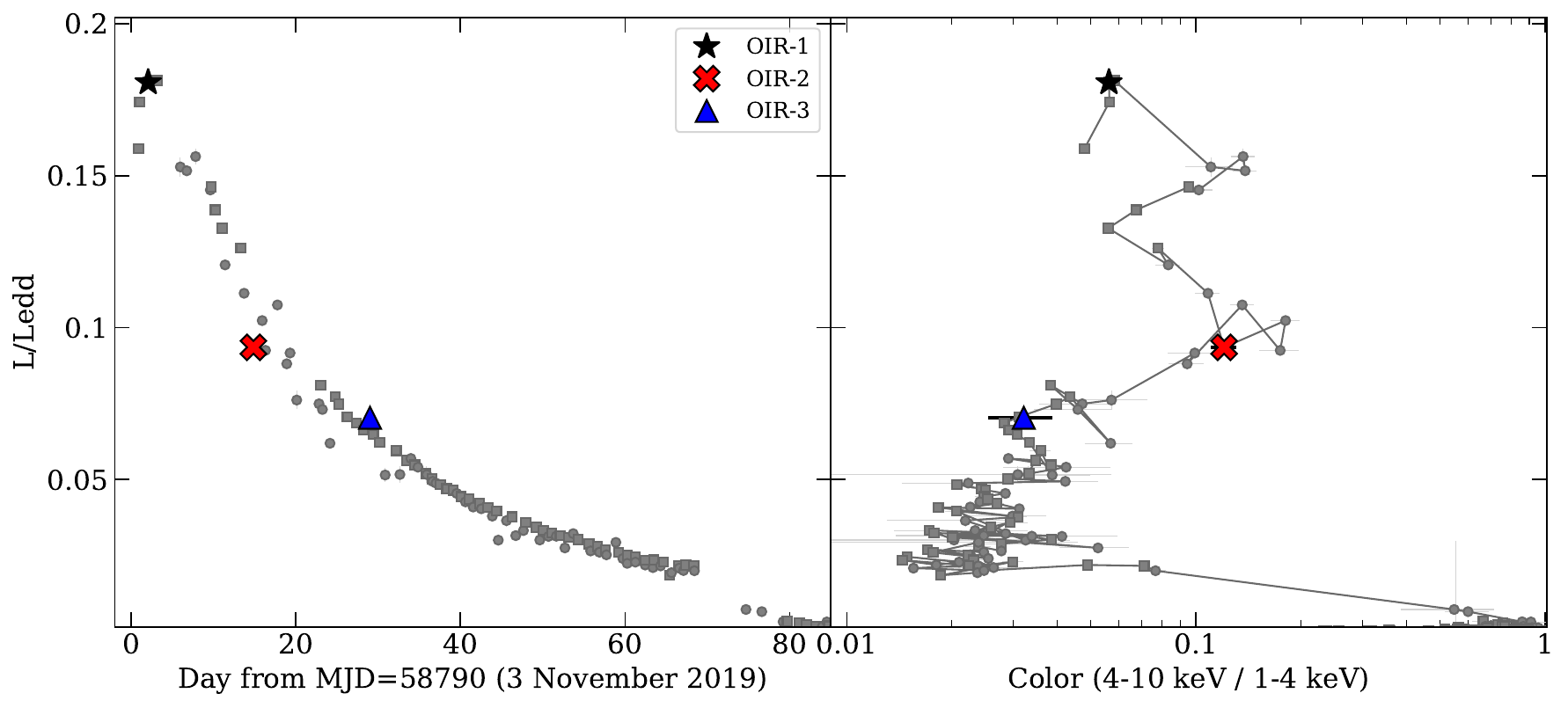}
\hspace{1cm}

\includegraphics[keepaspectratio,width=1.0\columnwidth, trim=2.0cm 0.0cm 2.0cm 0.0cm]{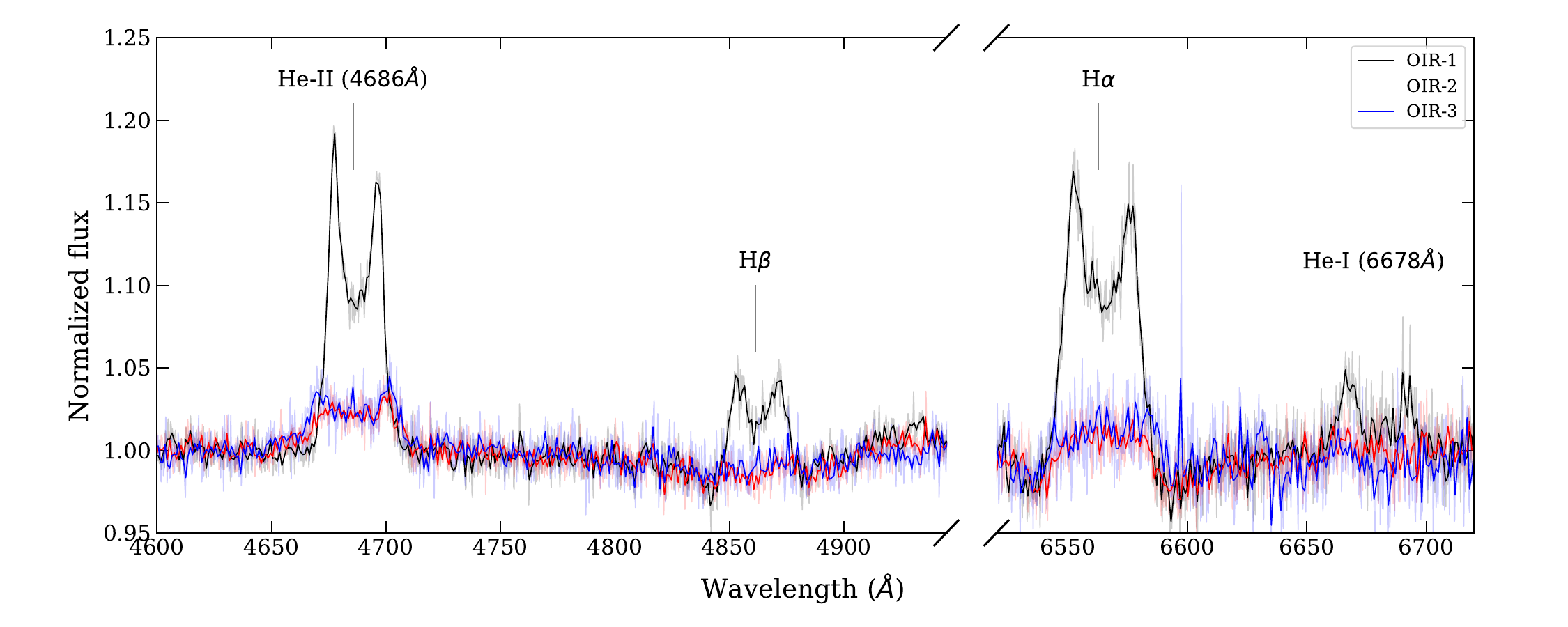}

\caption{\textbf{$\mid$ Spectral switch towards an irradiated accretion disk regime as the luminosity drops.}
\textit{Top panel}: X-ray light curve (left) and Hardness-Intensity Diagram (right) of \MAXI\ from \nicer\ and \swift\ (gray squares and circles, respectively). Error bars correspond to 90 per cent confidence intervals. The times of the three VLT epochs are indicated by a black star (OIR-1), a red cross (OIR-2), and a blue triangle (OIR-3). \textit{Bottom panel}: Trimmed, normalized optical spectra corresponding to the three VLT epochs. Epoch OIR-1 (black), taken at the peak of the outburst ($\sim$0.2~\ledd), shows strong double-peaked recombination lines of H and \hei. These lines become much weaker in epoch OIR-2 (red) and OIR-3 (blue), while the \heii\ emission remains relatively prominent, indicating a much hotter, irradiated accretion disk despite the lower X-ray luminosity ($\sim$0.1~\ledd).}
\label{fig:XrayLC_HID}
\end{figure}
}

\newcommand{\FigureHalphaPorb}{
\begin{figure*}[ht]
    \centering
    \includegraphics[keepaspectratio,width=1.0\columnwidth, trim=0.0cm 0.0cm 0.0cm 0.0cm]{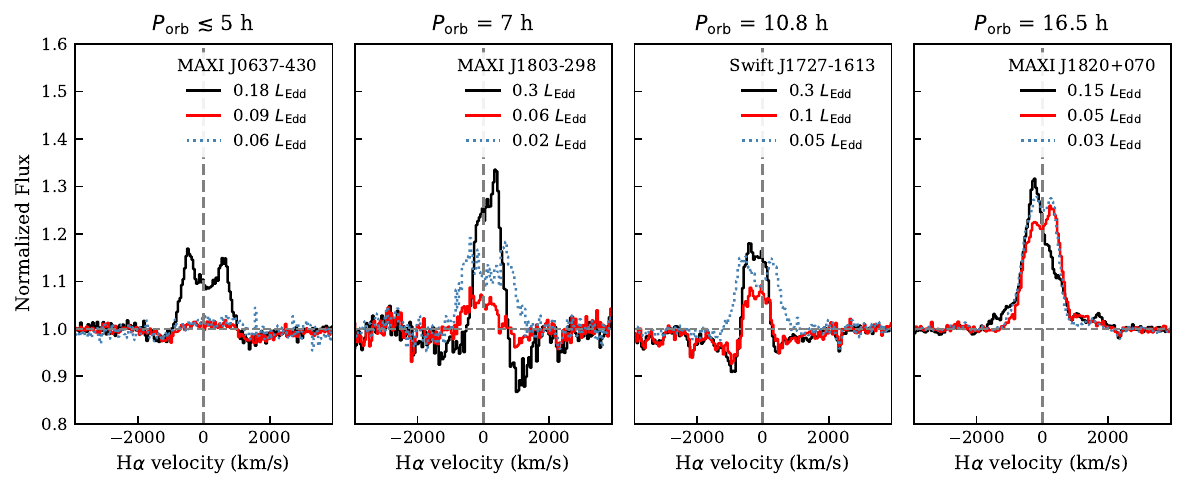}
    \caption{\textbf{$\mid$ Evolution of the H$\alpha$ line profile as a function of orbital period and luminosity.}
    Normalized H$\alpha$ profiles at three different luminosity stages during the soft-state decay of four black hole transients, ordered by increasing orbital period from left to right. The first panel shows the target of this study (\MAXI), which has an orbital period below 5~h, followed by MAXI~J1803$-$298 (7~h), Swift~J1727$-$1613 (10.8~h), and MAXI~J1820+070 (16.5~h). The H$\alpha$ line weakens or disappears at lower luminosities in the short-period systems, while it remains detectable in the longer-period sources. This behaviour is consistent with shielding of the outer disk at high luminosity in all cases. As the shielding vanishes at lower luminosities, small discs become fully ionised and suppress line formation, whereas extended discs retain partially neutral regions that continue to emit.
}
    \label{fig:HalphaEvol}
\end{figure*}
}
\newcommand{\FigureSketchver}{
\begin{figure}[H]
    \centering
    \includegraphics[keepaspectratio,width=1.0\columnwidth, trim=0.0cm 0.0cm 0.0cm 0.0cm]{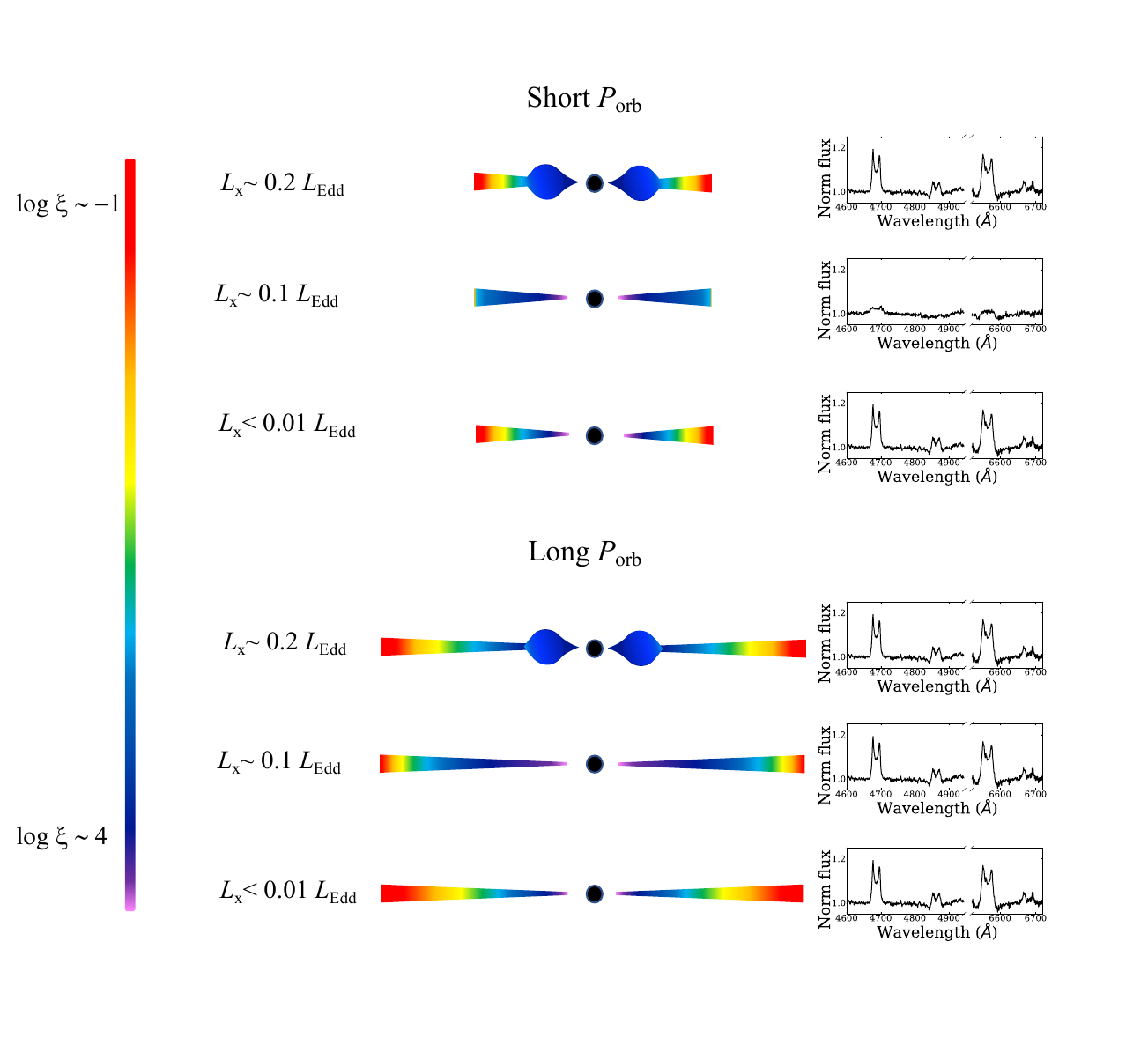}
    \caption{\textbf{$\mid$ Schematic view of the evolution of line formation and disk structure as a function of orbital period and luminosity.} Sketch illustrating the physical conditions that shape the OIR emission-line spectrum across luminosity and orbital period. The upper block corresponds to short-period systems ($\lesssim$ 11-15~hours) such as \MAXI, while the lower block represents systems with longer orbital periods ($\gtrsim$~11--15~hours). At high luminosities (top row), radiation pressure inflates the inner disk, shielding the outer regions and enabling the formation of strong recombination lines. As the luminosity drops (moving downward), the disk returns to a thin configuration, the outer regions become over-ionized, and the emission lines weaken or vanish in short-period systems. In long-period systems, the outer disk lies much farther from the X-ray source, allowing line formation to persist even under high irradiation due to lower temperatures at larger radii.}
    \label{fig:sketch}
\end{figure}
}

\newcommand{\FigureOIRCont}{
\begin{figure}[h]
\centering
\includegraphics[keepaspectratio,width=1.0\columnwidth, trim=0.0cm 0.0cm 0.0cm 0.0cm]{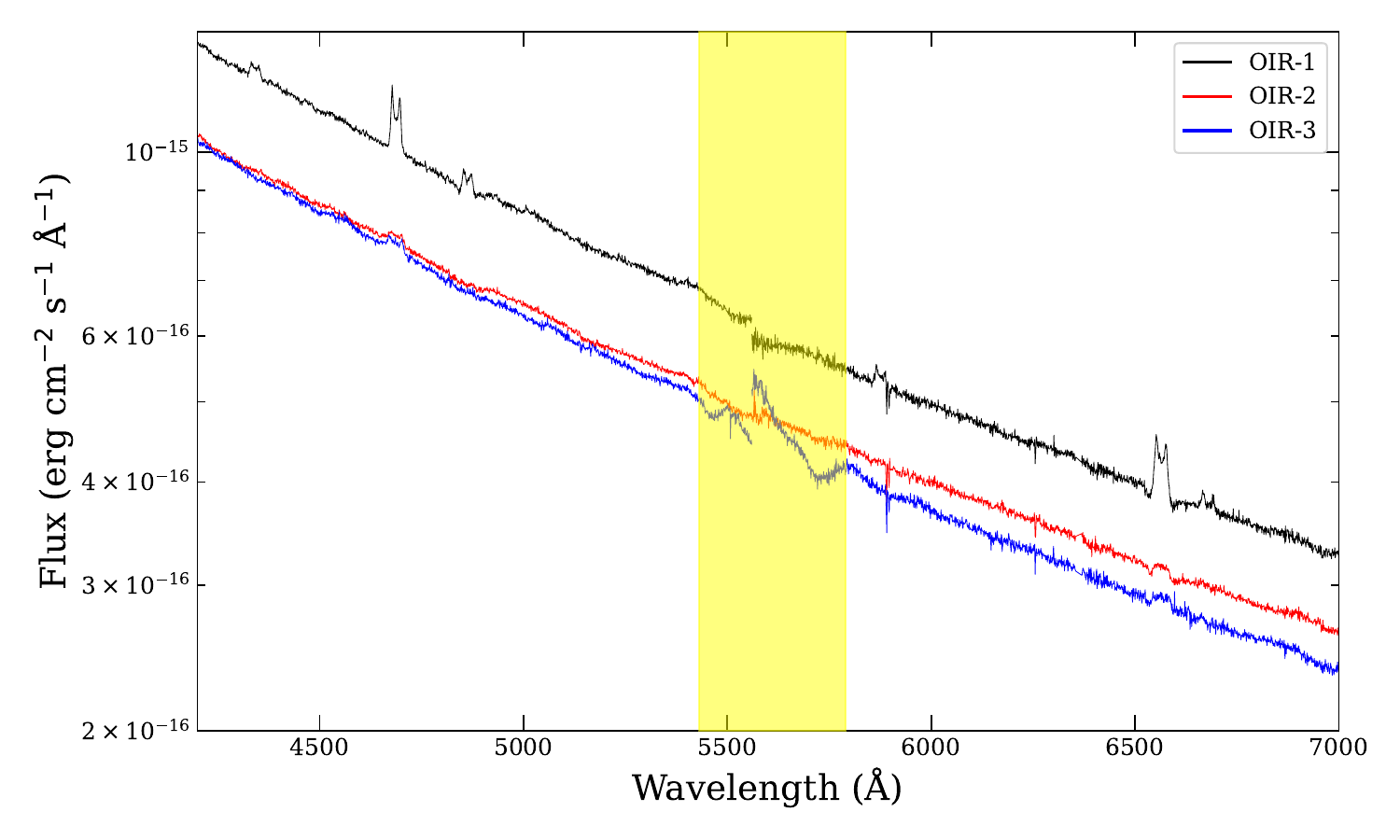}

\caption{\textbf{$\mid$ Flux-calibrated optical spectra during the outburst decay.}
Flux-calibrated optical spectra of \MAXI\ corresponding to the three OIR epochs shown in Fig.~\ref{fig:XrayLC_HID}. Epoch OIR-1 (black) displays strong double-peaked recombination lines of H and He, superimposed on a bright continuum. These features weaken significantly in the epoch OIR-2 (red) and are nearly absent by OIR-3 (blue) consistent with a hotter and more ionised accretion disk. The overall continuum level drops steadily across the three epochs, following the X-ray luminosity decay, in agreement with the expected evolution of an X-ray irradiated disk. The shaded yellow region marks the wavelength gap between the UVB and VIS arms of the X-shooter spectrograph.}
\label{fig:oir_continuum}
\end{figure}
}

\newcommand{\ThreeNustar}{
\begin{figure}[ht]
\centering

\includegraphics[width=0.8\textwidth]{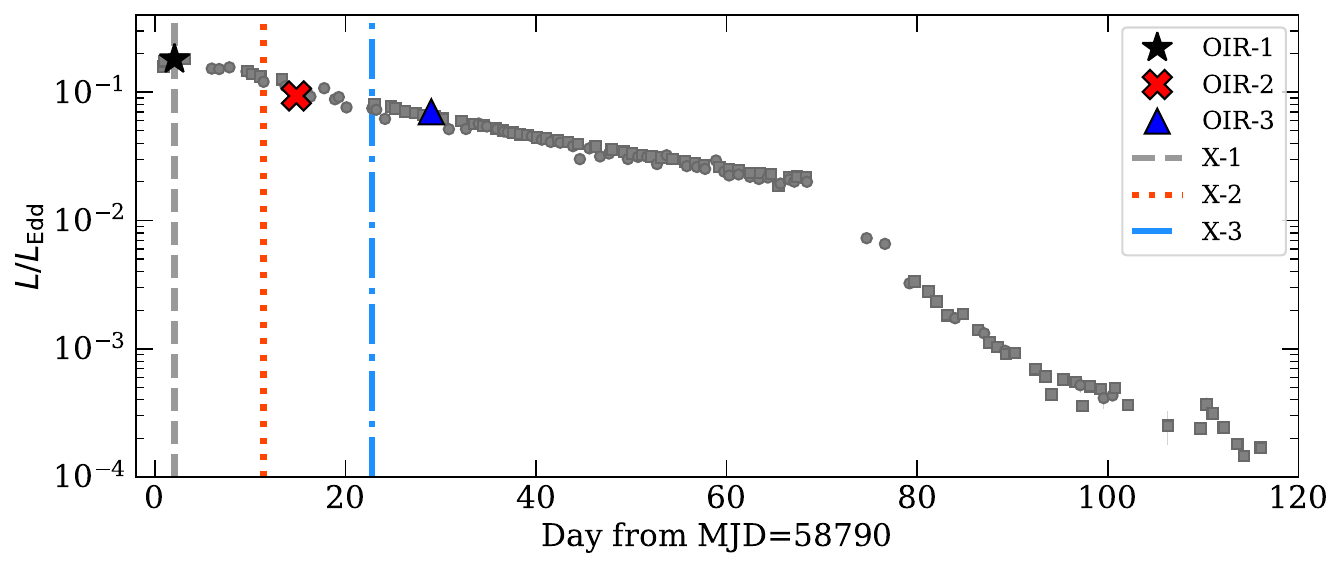} \\[1ex]

\begin{tabular}{cc}
\includegraphics[width=0.5\textwidth]{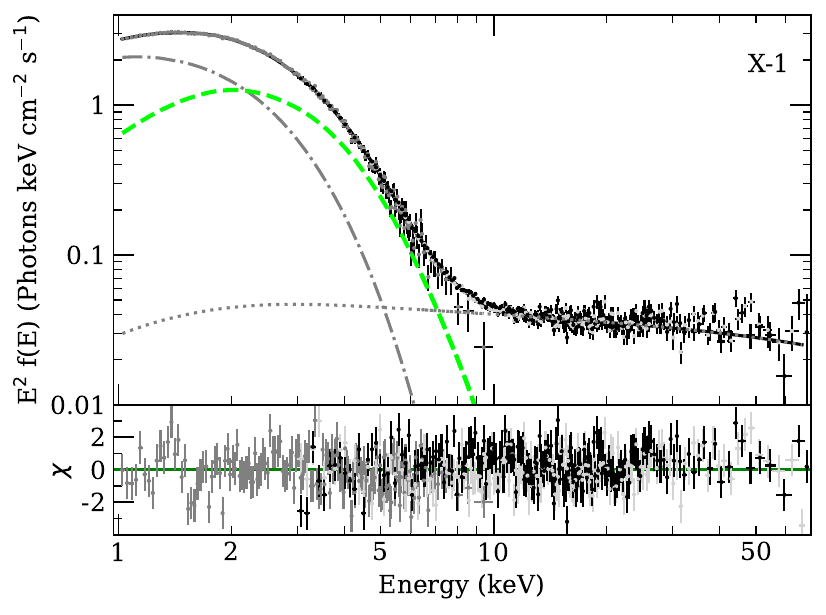} &
\includegraphics[width=0.5\textwidth]{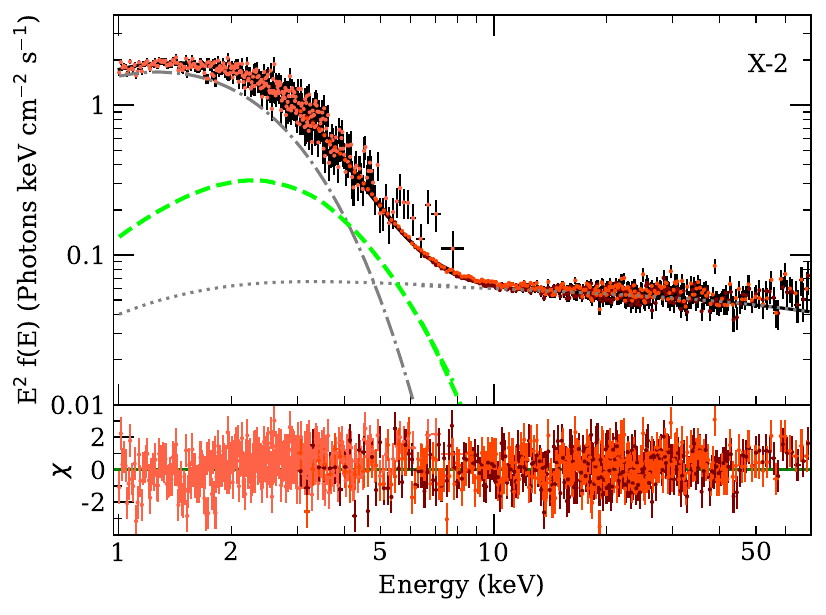} \\
\end{tabular} \\[1ex]

\includegraphics[width=0.5\textwidth]{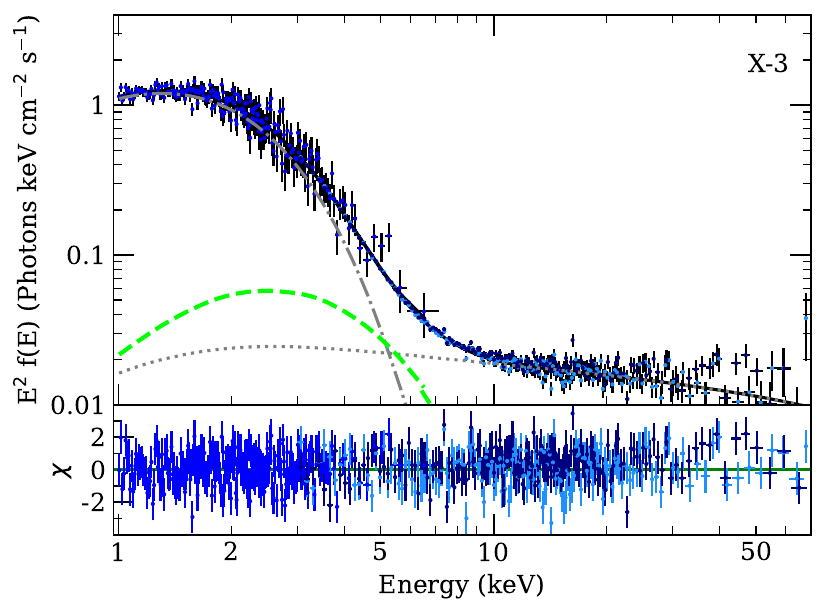}

\caption{\textbf{Broad-band X-ray spectral evolution during the outburst decay.} \textit{Top panel}: X-ray light curve of MAXI~J0637$-$430 from \nicer\ and \swift\ (grey points), showing the evolution of the Eddington-scaled luminosity throughout the outburst. The three OIR epochs of VLT spectroscopy are marked with colored symbols (black star, red cross, and blue triangle). Vertical grey dashed, orange dotted, and blue dash-dotted lines indicate the times of the three broad-band X-ray observations, which combine \nicer/\swift\ with simultaneous \nustar\ data.
\textit{Bottom panels}: Unfolded X-ray spectra for the three broad-band X-ray epochs combining \nicer\ and \nustar\ data for epoch X-1 (grey) , and \swift\ and \nustar\ data for epochs X-2 and X-3 (orange and blue, respectively). The best fit obtained with the model \texttt{Const*Tbabs(diskpbb+Nthcomp+relxillNS)} is shown as a solid line, with the thermal disk (dash-dotted), Comptonised (dotted), and reflection (green dashed) components shown individually. Residuals (in units of $\sigma$) are plotted below each spectrum. All uncertainties are given at the 90 per cent confidence level.}
\label{fig:threeNustar}
\end{figure}
}

\newcommand{\FigureNUSTAR}{
\begin{figure*}$\phantom{ht}$
\centering
\includegraphics[keepaspectratio,width=15cm, trim=0.0cm 0.0cm 0.0cm 0.0cm]{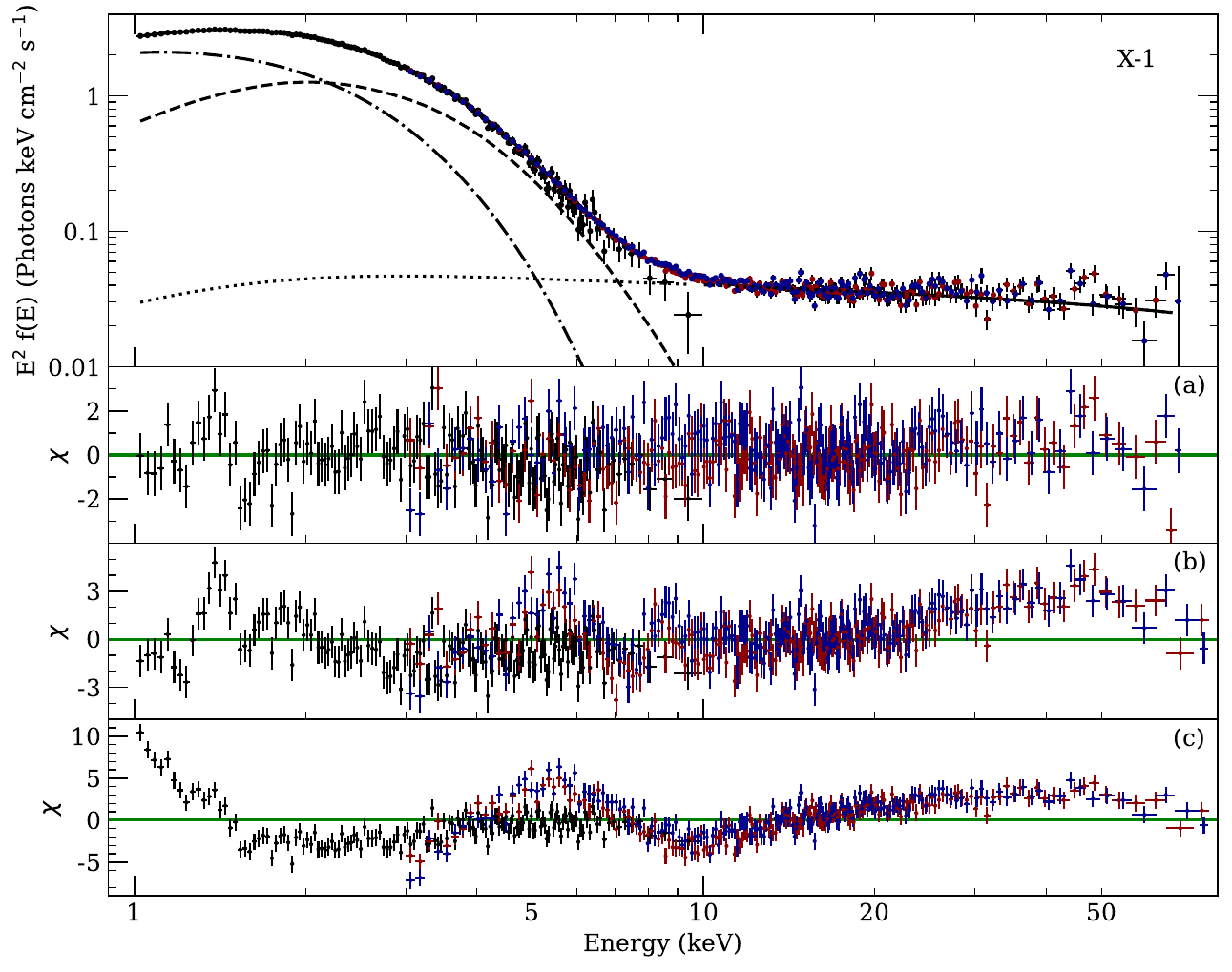}
\caption{\textbf{Spectral residuals of alternative models for epoch X-1.}
Unfolded 1–70~keV X-ray spectrum of \MAXI\ from epoch X-1 (\nicer\ in black, and \nustar\ in dark red and dark blue for FPMA and FPMB, respectively). The solid line shows the best fit obtained with \texttt{Const*Tbabs(diskpbb+Nthcomp+relxillNS)}, with the individual thermal, Comptonized, and reflection components shown as dash–dotted, dashed, and dotted lines, respectively. Residuals (in units of $\sigma$) are shown in the subpanels for fits with: (a) \texttt{Const*Tbabs(diskpbb+Nthcomp+relxillNS)} (final model), (b) \texttt{Const*Tbabs(diskbb+relxillCp)}, and (c) \texttt{Const*Tbabs(diskbb+Nthcomp)}.}
\label{fig:nustarspectra}
\end{figure*}
}
\newcommand{\FigureVLTUnoDos}{
\begin{figure}[ht]
\centering
\includegraphics[keepaspectratio,width=0.9\columnwidth, trim=0.0cm 0.0cm 0.0cm 0.0cm]{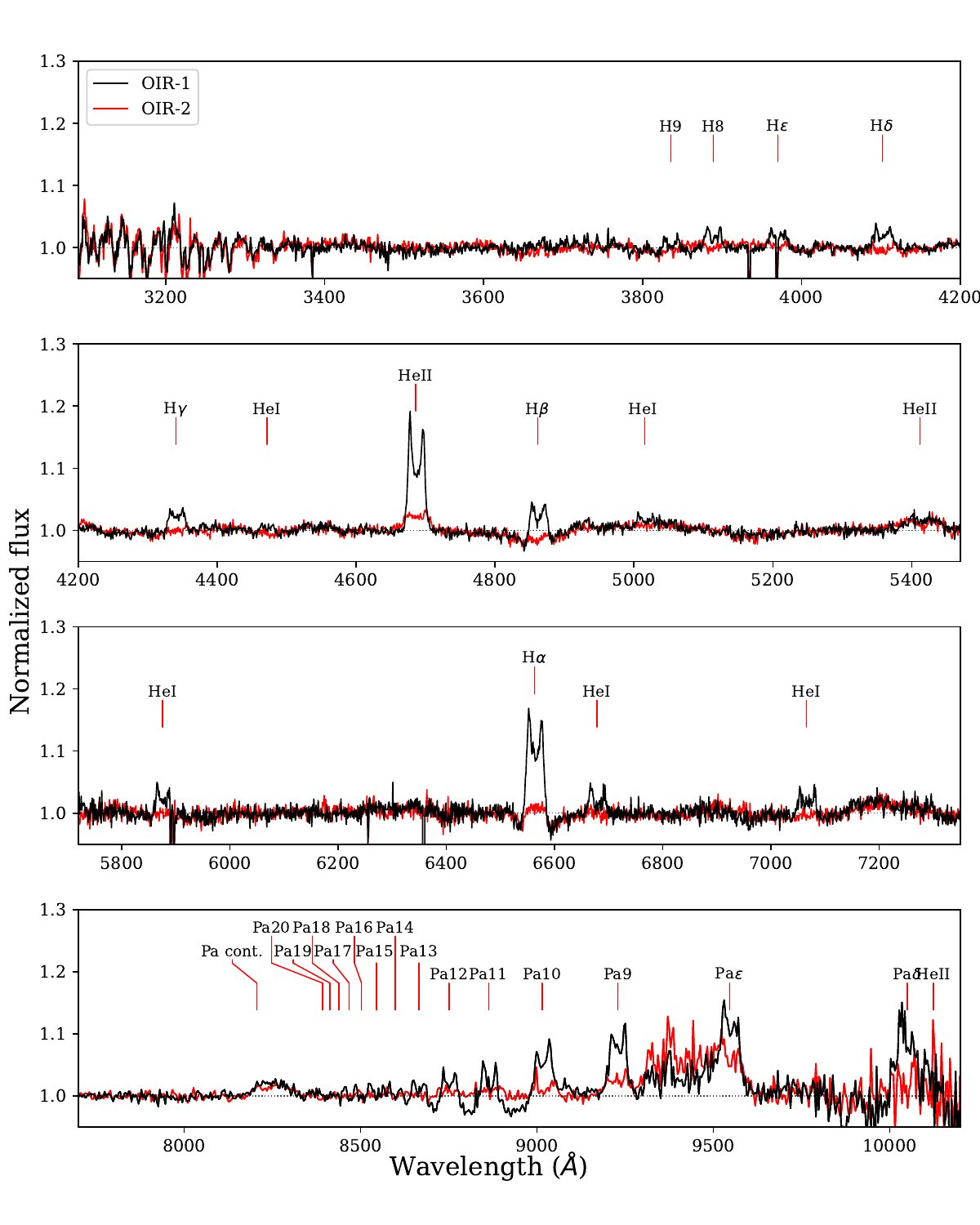}
\caption{\textbf{$\mid$ Full optical and near-infrared spectra during two epochs of the outburst decay.} Normalised VLT/X-shooter spectra of epoch OIR-1 (black) and OIR-2 (red) of \MAXI, showing the full optical and near-infrared coverage. Major emission line features are marked, including Balmer and Paschen series lines, as well as prominent He\,\textsc{i} and He\,\textsc{ii} transitions. During epoch OIR-1, strong double-peaked Balmer and He\,\textsc{i} emission lines are clearly visible, characteristic of an optically thick, low-ionisation accretion disc. By OIR-2, these features have significantly weakened or disappeared, consistent with an increase in ionisation conditions in the outer disc.
}
\label{fig:VLT_1_2}
\end{figure}
}

\newcommand{\FigureVLTUnoTres}{
\begin{figure}[ht]
\centering
\includegraphics[keepaspectratio,width=0.9\columnwidth, trim=0.0cm 0.0cm 0.0cm 0.0cm]{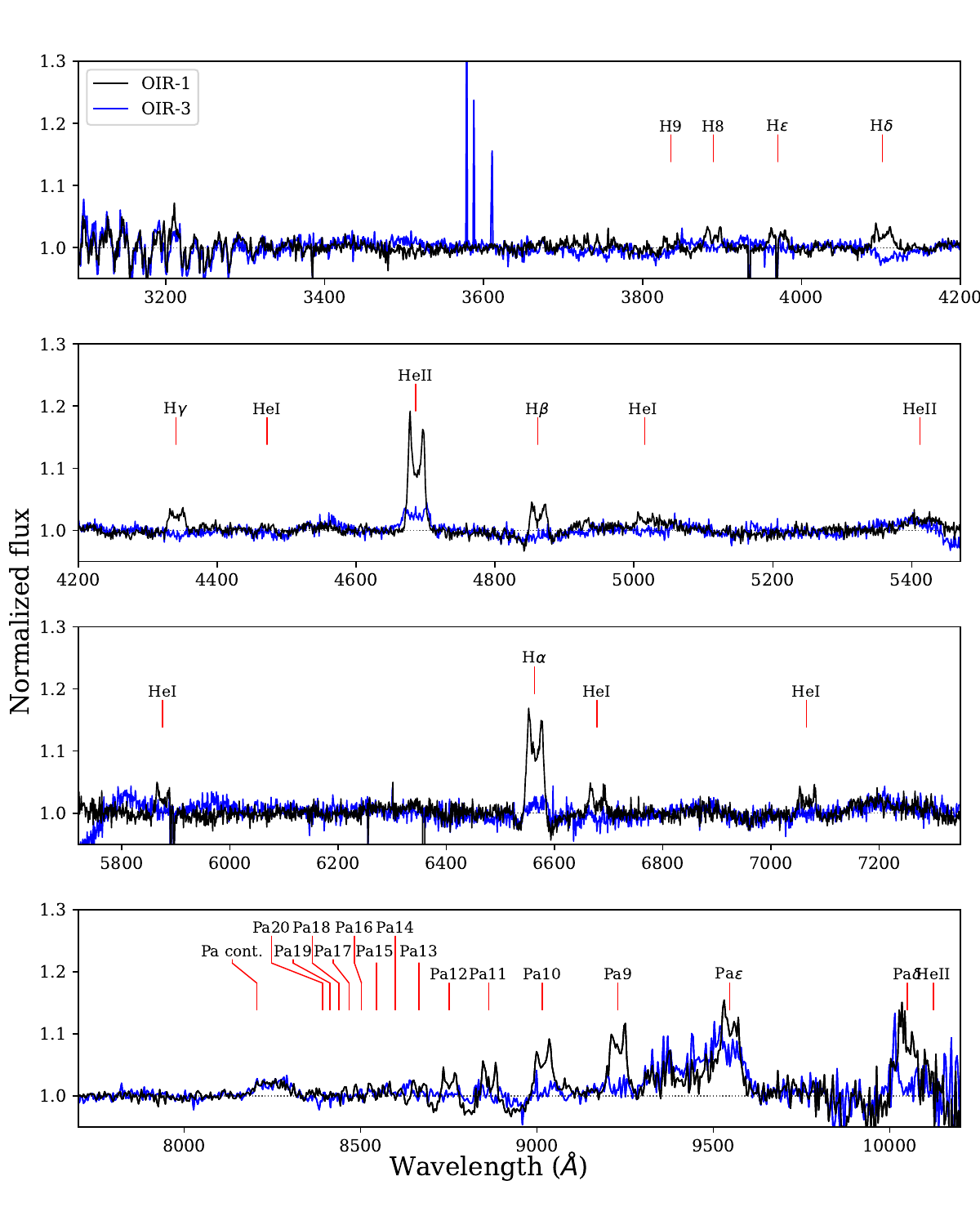}
\caption{\textbf{$\mid$ Full optical and near-infrared spectra during two epochs of the outburst decay.} Normalised VLT/X-shooter spectra of epoch OIR-1 (black) and OIR-3 (blue) of \MAXI, showing the full optical and near-infrared coverage. As in the comparison between epochs OIR-1 and OIR-2 (Fig.\ref{fig:VLT_1_2}), major emission lines such as those of the Balmer and Paschen series, as well as He\,\textsc{i} and He\,\textsc{ii}, are marked. Strong Balmer and He\,\textsc{i} lines observed during OIR-1 have further weakened or disappeared by epoch OIR-3, consistent with a continued increase in the ionisation of the outer disc.
}
\label{fig:VLT_1_3}
\end{figure}
}

\newcommand{\FigureEW}{
\begin{figure*}$\phantom{ht}$
\centering
\includegraphics[keepaspectratio,width=15cm, trim=0.0cm 0.0cm 0.0cm 0.0cm]{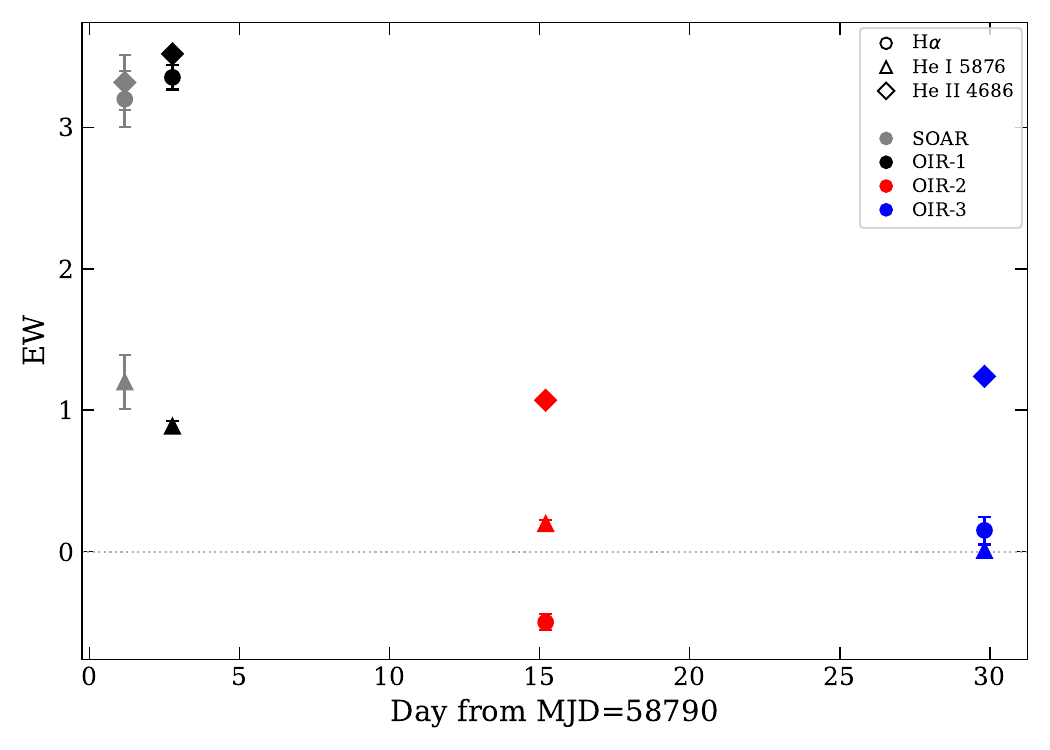}
\caption{\textbf{Evolution of optical emission line strengths during the outburst decay}. Equivalent widths (EW) of three representative optical emission lines -- \ha\ (circles), \hei\ $\lambda$5876 (triangles), and \heii\ $\lambda$4686 (diamonds) -- as a function of time (MJD). Different colours mark the spectroscopic epochs: SOAR (grey), OIR-1 (black), OIR-2 (red), and OIR-3 (blue). The EW of all lines decreases after OIR-1, with Balmer and \hei\ lines fading to almost undetectable levels. \heii\ $\lambda$4686 also declines, but remains visible throughout, indicating a much hotter, ionised, irradiated accretion disc. Error bars correspond to 1$\sigma$ uncertainties. The negative \ha\ EW at OIR-2 likely reflects noise fluctuations in the absence of line emission, compounded by broad underlying Balmer absorption commonly seen in LMXB spectra \cite{MataSanchez2026}.}
\label{fig:EW}
\end{figure*}
}

\newcommand{\FigureSOAR}{
\begin{figure*}$\phantom{ht}$
\centering
\includegraphics[keepaspectratio, width=1.0\columnwidth, trim=0.0cm 0.0cm 0.0cm 0.0cm]{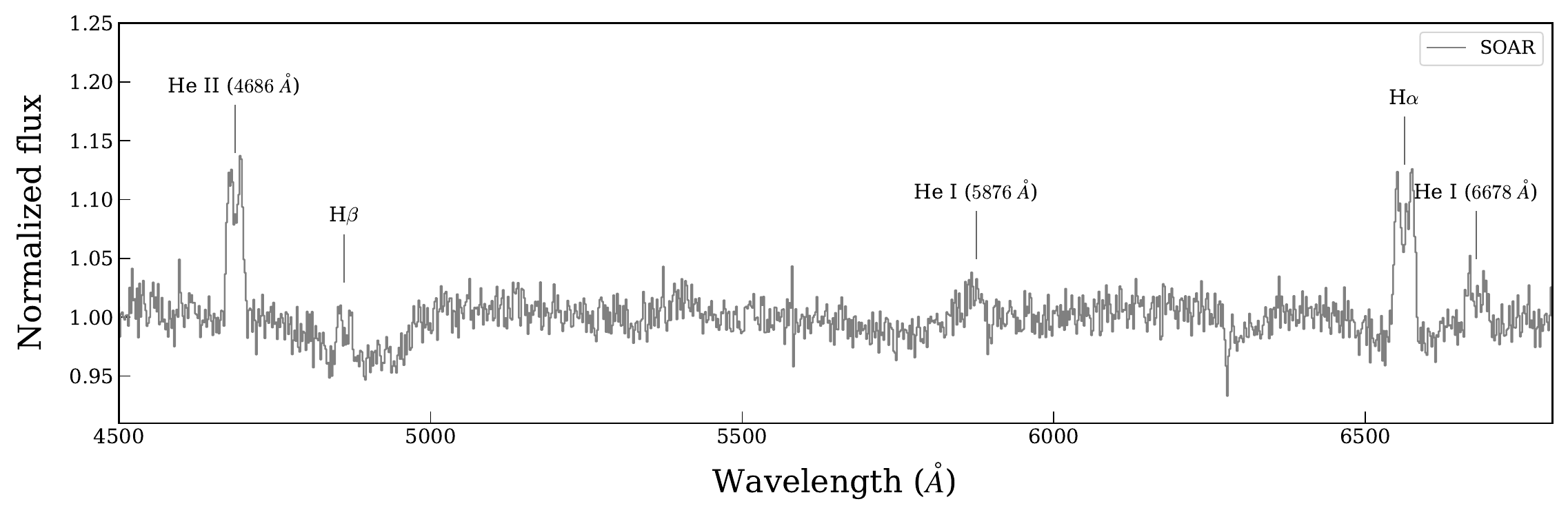}
\caption{\textbf{Optical spectrum obtained with SOAR one day before our first VLT epoch.} Normalised optical spectrum of \MAXI\ taken with the Goodman Spectrograph on the SOAR telescope on 2019 Nov~4, one day before epoch OIR-1 \cite{Strader2019}. Prominent double-peaked emission lines from H$\alpha$, H$\beta$, \hei~$\lambda6678$ and \heii~$\lambda4686$ are clearly visible. These features, also observed in our OIR-1 spectrum, confirm the presence of a bright, low-ionisation accretion disk early in the outburst.}
\label{fig:SOAR}
\end{figure*}
}

%% file: z2_Tables.tex
\newcommand{\TableFitNuSTARNEW}{
\begin{table*}
\centering
\begin{threeparttable}

\caption{Best-fit parameters for the \texttt{Const*Tbabs(diskpbb+Nthcomp+relxillNS)} model at the three broad-band X-ray epochs. Epoch X-1 corresponds to the joint \nicer+\nustar\ spectrum, while X-2 and X-3 correspond to the joint \swift+\nustar\ spectra.}

\begin{tabular}{lccc}
\hline\hline
 & \textbf{Epoch X-1} & \textbf{Epoch X-2} & \textbf{Epoch X-3} \\
\hline
\texttt{TBABS} &   &   & \\
\Nh\ ($\times10^{-22}$ \nh)                             &  \multicolumn{3}{c}{0.04 (fixed)} \\
\hline
\texttt{DISKPBB} &   &   & \\
$kT_{\rm in}$ (keV)      & $0.522^{+0.006}_{-0.017}$        & $0.523^{+0.005}_{-0.014}$       & $0.506^{+0.005}_{-0.003}$ \\
$p$                       & $0.58 \pm 0.01$                 & $0.73\pm 0.04$       & $0.78^{+0.01}_{-0.06}$ \\
$N_{\rm diskpbb}$        & $2392^{+330}_{-241}$             & $3021^{+514}_{-430}$           & $3172^{+42}_{-540}$ \\
\hline
\texttt{NTHCOMP} &   &   & \\
$\Gamma$                 & $2.14\pm 0.02$                   & $2.09\pm 0.02$                  & $2.22 \pm 0.03$ \\
$kT_{\rm e}$ (keV)       & $>81$                            & $>145$                            & $>45$ \\
$kT_{\rm seed}$ (keV)    &  \multicolumn{3}{c}{(linked to $kT_{\rm in}$)} \\
$N_{\rm nthcomp}$        & $0.032 \pm 0.001$                & $0.044 \pm 0.002$               & $0.018 \pm 0.001$ \\
\hline
\texttt{RELXILLNS} &   &   & \\
$a_*$                             &  \multicolumn{3}{c}{$0.58^{+0.08}_{-0.11}$ (tied)} \\
$i$ (deg)$^{\mathrm{c}}$                        &  \multicolumn{3}{c}{$69.98^{+0.01}_{-2.17}$ (tied)} \\
$kT_{\rm bb}$ (keV)      & $0.523 \pm 0.002$                & $0.55 \pm 0.001$                 & $0.66 \pm 0.02$ \\
$\log\xi$                & $4.4^{+0.2}_{-0.3}$           & $4.4^{+0.2}_{-0.4}$          & $4.5^{+0.1}_{-0.5}$ \\
$A_{\rm Fe}$ (solar)              & \multicolumn{3}{c}{$1.4 \pm 0.2$ (tied)} \\
$\log N$ (cm$^{-3}$)      & $16.03 \pm 0.01$          & $19.00^{b}$         & $17 \pm 2$ \\
$N_{\rm relxillNS}$ ($\times10^{-3}$) & $6.2^{+0.4}_{-0.2}$           & $2.1^{+0.3}_{-0.1}$         & $0.30^{+0.03}_{-0.04}$ \\
\hline
$F_{\rm x}$$^{\mathrm{a}}$($\times10^{-9}$ erg cm$^{-2}$ s$^{-1}$) & $8.76 \pm 0.02$ & $5.38 \pm 0.05$    & $3.28 \pm 0.05$ \\
Diskpbb fraction         &    63                &    73        &  91   \\
RelxillNS fraction       & 33                   & 18          &  4  \\
Nthcomp fraction        & 4                 &    9           &    5   \\
$\chi^2$/dof  &  \multicolumn{3}{c}{$4532.05/4233$ = 1.07} \\
\hline
\end{tabular}
\begin{tablenotes}
\item[a]{Unabsorbed 0.5--100 keV X-ray flux in units of $\times10^{-9}$ erg cm$^{-2}$ s$^{-1}$. We derived the unabsorbed fluxes and uncertainties in the 0.5--100 keV range using \texttt{CFLUX} and the best-fitting model. }
\item[b]{Parameter pegged upper bound during fit}
\item[c]{An upper limit of 70 deg was imposed.}
\item[]{\textit{Note}: Uncertainties are quoted at the 90\% confidence level and calculated from Markov Chain Monte Carlo (MCMC) of chain length $10^6$}
\end{tablenotes}
\label{tab:fit_three_epochs_Nustar}
\end{threeparttable}
\end{table*}
}

%% file: definitions.tex
\newcommand{\hei}{He~{\sc i}}
\newcommand{\heii}{He~{\sc ii}}
\newcommand{\ha}{H$\alpha$}

\def\chired{$\chi^{2}_{\nu}$}

\newcommand{\ledd}{$L_\mathrm{Edd}$}
\newcommand{\Nh}{$N_{\rm H}$}

\newcommand{\flux}{\mathrm{erg~cm}^{-2}~\mathrm{s}^{-1}}

\newcommand{\nh}{$\mathrm{cm}^{-2}$}

\newcommand{\swift}{\textit{Swift}}

\newcommand{\nustar}{\textit{NuSTAR}}

\newcommand{\nicer}{\textit{NICER}}